\documentclass[preprint, aps, amssymb, prl, superscriptaddress]{revtex4}  
\usepackage[draft]{hyperref} 
\usepackage{amsmath}
\usepackage{amssymb}
\usepackage{amsthm}
\usepackage{wasysym}
\usepackage{graphicx}

\begin{document}
\title{Competition between electronic Kerr and free carrier effects in an ultimate-fast optically switched semiconductor microcavity}

\author{Emre Y\"uce}\email{e.yuce@utwente.nl: www.photonicbandgaps.com}
\affiliation{Complex Photonic Systems (COPS), MESA+ Institute for Nanotechnology, University of Twente, The Netherlands}
\author{Georgios Ctistis}
\affiliation{Complex Photonic Systems (COPS), MESA+ Institute for Nanotechnology, University of Twente, The Netherlands}
\author{Julien Claudon}
\affiliation{CEA-CNRS-UJF ``Nanophysics and Semiconductors" joint laboratory, CEA/INAC/SP2M, 17 rue des Martyrs, 38054 Grenoble Cedex 9 France}
\author{Emmanuel Dupuy}
\affiliation{CEA-CNRS-UJF ``Nanophysics and Semiconductors" joint laboratory, CEA/INAC/SP2M, 17 rue des Martyrs, 38054 Grenoble Cedex 9 France}
\author{Klaus J. Boller}
\affiliation{Laser Physics $\&$ Nonlinear Optics Group (LPNO), MESA+ Institute for Nanotechnology, University of Twente, The Netherlands}
\author{Jean-Michel G\'erard}\email{jean-michel.gerard@cea.fr}
\affiliation{CEA-CNRS-UJF ``Nanophysics and Semiconductors" joint laboratory, CEA/INAC/SP2M, 17 rue des Martyrs, 38054 Grenoble Cedex 9 France}
\author{Willem L. Vos}\email{W.L.Vos@tnw.utwente.nl}
\affiliation{Complex Photonic Systems (COPS), MESA+ Institute for Nanotechnology, University of Twente, The Netherlands}

\begin{abstract}We have performed ultrafast pump-probe experiments on a GaAs-AlAs microcavity with a resonance near 1300 nm in the ``original" telecom band. We concentrate on ultimate-fast optical switching of the cavity resonance that is measured as a function of pump-pulse energy. We observe that at low pump-pulse energies the switching of the cavity resonance is governed by the instantaneous electronic Kerr effect and is achieved within 300 fs. At high pump-pulse energies the index change induced by free carriers generated in the GaAs start to compete with the electronic Kerr effect and reduce the resonance frequency shift. We have developed an analytic model which predicts this competition in agreement with the experimental data. To this end we derive the nondegenerate two- and three-photon absorption coefficients for GaAs. Our model includes a new term in the intensity-dependent refractive index that considers the effect of the probe pulse intensity, which is resonantly enhanced by the cavity. We calculate the effect of the resonantly enhanced probe light on the refractive index change induced by the electronic Kerr effect for cavities with different quality factors. By exploiting the linear regime where only the electronic Kerr effect is observed, we manage to retrieve the nondegenerate third order nonlinear susceptibility $\chi^{(3)}$ for GaAs from the cavity resonance shift as a function of pump-pulse energy. 
\end{abstract}

\maketitle

\section{Introduction}

Semiconductor cavities have attracted considerable attention in recent years due to their ability to store light for a given amount of time in a small volume \cite{vahala.2003.nature}. This key issue of cavities stimulated a large amount of experiments for increasing the nonlinear interaction of photons and to understand and exploit cavity quantum electrodynamics (cQED) effects \cite{abad.2007.oe, allison.2011.prl, imamoglu.2012.nat.ph, gerard.2003.t.app.phys, reithmaier.2008.sem.sci.tech}. The dynamic manipulation of these systems, especially of combined cavity emitter systems is thereby of major interest \cite{vos.2002.prb, reithmaier.2004.nature, vuckovic.2012.prl}. All-optical switching of cavities gains momentum since it enables the dynamic control of the capture and release of photons on sub-picosecond timescales \cite{georgios.2011.apl}. Moreover, ultrafast change of the optical properties of a cavity prevails to frequency conversion through adiabatic \cite{notomi.2009.prl} and not-adiabatic processes \cite{harding.2012.josab}.

The optical properties of cavities can be altered by changing the refractive index of the constituent material. The refractive index of a semiconductor cavity can be switched via the excitation of free-carriers in the semiconductor \cite{lipson.2004.nature, wong.2009.apl, yuce.2009.ieee, notomi.2010.nat.ph, jewell.1989.apl, rivera.1994.apl}. However, the switching speed in such schemes is material dependent and limited by the recombination dynamics of the excited carriers. On the other hand, the refractive index of a semiconductor cavity can also be changed with the electronic Kerr effect. The electronic Kerr effect is, in terms of speed, the ultimate way for ultrafast switching due to its material independent and instantaneous response nature \cite{georgios.2011.apl}. Yet, the excitation of relatively slow free carriers has to be avoided to accomplish a positive refractive index change with the electronic Kerr effect, since free carriers lead to an opposite change of refractive index \cite{jalali.2004.oe, johnson.2006.oe, harding.2009.josab}. The main challenge is therefore to find a range of parameters where solely the Kerr effect controls the optical properties of the cavity.

In this work we employ the instantaneous electronic Kerr effect to switch the resonance frequency of a semiconductor planar microcavity in the original telecom band within 300 fs as a function of pump-pulse energy. Using two light sources that provide pump and probe pulses we observe and analyse the competition of free carrier induced index changes and the electronic Kerr effect in a switched cavity. We have developed an analytical model which predicts this competition in agreement with our experimental data. Our model is developed for nondegenerate light sources and thereby can explain the effect of the cavity enhancement and the intensity of each source in the switching of a cavity resonance. 

\section{Sample and Experimental Setup}

\begin{figure}[H]
  \begin{center}
  \includegraphics[width=8.2 cm]{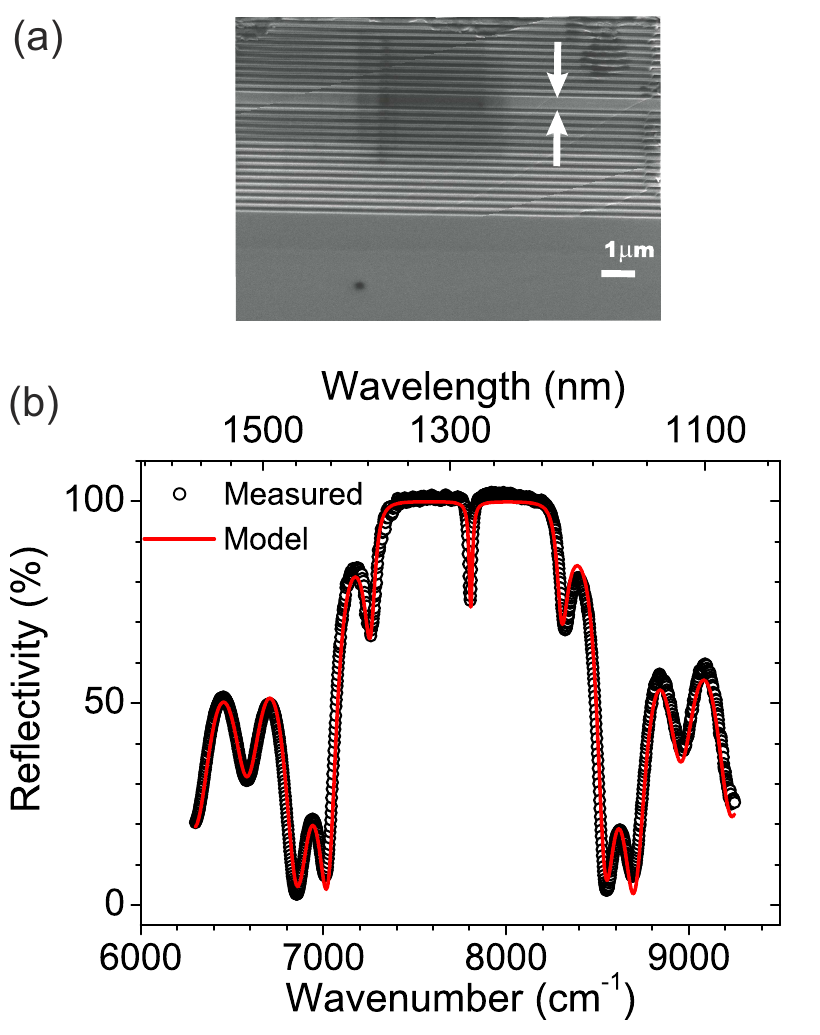}
  \caption{(color online) \emph{(a) Scanning electron micrograph of our microcavity. The GaAs $\lambda$-layer is indicated with white arrows and is sandwiched between two GaAs-AlAs Bragg stacks. The GaAs substrate is visible at the bottom. The GaAs layers appear dark grey, while the AlAs layers appear light grey. (b) Measured (black symbols) and calculated (red line) reflectivity spectra of the microcavity. The stopband of the Bragg stacks extends from $7072 \ \rm{cm^{-1}}$ to $8498 \ \rm{cm^{-1}}$. Fabry-P\'erot fringes are visible on both sides of stop band. Within the stopband a narrow trough at $7794.2 \ cm^{-1}$ ($1282 \ nm$) indicates the cavity resonance. From the linewidth ($\Delta \omega=20  \pm 3 \ \rm{cm^{-1}}$, full width at half maximum) of the cavity resonance we derive a quality factor $Q=390 \pm 60$ corresponding to a cavity storage time of $\tau_{cav}=0.3 \pm 0.045 \ \rm{ps}$. The calculations are performed with a transfer matrix model.}}
\label{yuce-f1}
  \end{center}
\end{figure}

Our experiments are performed on a planar microcavity grown by means of molecular-beam epitaxy. The cavity resonance is designed to occur at $\lambda_0=1280 \pm 5 \ \rm{nm}$ in the Original (\emph{O}) telecom band. Fig. \ref{yuce-f1} (a) shows a scanning electron micrograph of the sample. The sample consists of a GaAs $\lambda$-layer ($d=376 \ \rm{nm}$) sandwiched between two Bragg stacks made of 7 and 19 pairs of $\lambda/4$-thick layers of nominally pure GaAs ($d_{GaAs}=94 \ \rm{nm}$) and AlAs ($d_{AlAs}= 110 \ \rm{nm}$), respectively and positioned on a GaAs wafer. The storage time of the probe photons in the cavity is deliberately reduced by decreasing the reflectivity of the top mirror of the cavity. This leads to faster switching rates while at the same time reducing free carrier excitation due to a reduced field enhancement in the cavity. 

\begin{figure}
  \begin{center}
  \includegraphics[width=8.2 cm]{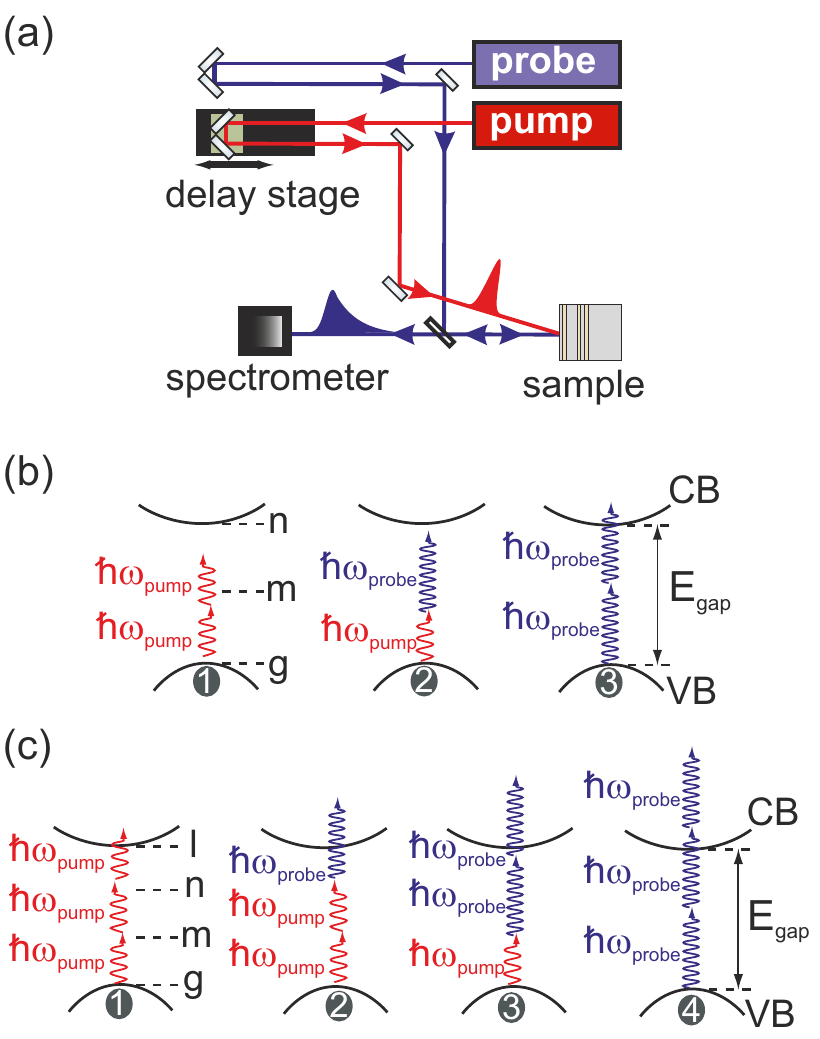}
  \caption{(color online) \emph{(a) Schematic of the setup. The probe beam path is shown in blue, the pump beam path in red. The time delay between the pump and the probe pulses is adjusted through a delay stage. The reflected signal from the cavity is spectrally resolved and detected with a spectrometer. The frequency of the probe beam is resonant with the cavity and the bandwidth of the probe beam is broader than the cavity linewidth. (b)Schematic energy diagram for the two-photon carrier excitation processes possible in our experiment (see Eq. \ref{4twophotondensity}). Two-photon absorption is largely suppressed by the judicious tuning of the pump and the probe frequencies relative to the semiconductor bandgap energy (c) Schematic energy diagrams for all possible three-photon processes in the experiment that may result in free carrier generation (see Eq. \ref{5threephotondensity}).}}
\label{yuce-f2}
  \end{center}
\end{figure}

Figure  \ref{yuce-f1} (b) shows the measured and the calculated reflectivity spectrum of the microcavity. The reflectivity spectrum of the cavity is measured with a setup consisting of a supercontinuum broad-band white-light source and a Fourier-transform interferometer with a resolution of $0.5 \ \rm{cm^{-1}}$ (BioRad FTS6000). It can be seen that the stopband of the Bragg stack extends from $7072 \ \rm{cm^{-1}}$ to $8498 \ \rm{cm^{-1}}$ ($1414 \ \rm{nm} \ to \ 1177 \ \rm{nm}$).  On both sides of the stopband Fabry-P\'erot fringes are visible due to interference of the light reflected from the front and the back surfaces of the sample. Inside the stopband, a narrow trough indicates the cavity resonance at $\omega_{res}= 7794.2 \ \rm{cm^{-1}}$ ($\lambda_{res}=1283.01 \ \rm{nm}$). The resonance frequency of the switched cavity is determined by fitting a Lorentz function to the resonance trough over a limited frequency range. The relatively high reflectivity of the resonance minimum ($R_{trough}=80\ \%$) is a result of the asymmetric cavity design. From the linewidth ($\Delta \omega=20  \pm 3 \ \rm{cm^{-1}}$, full width at half maximum) of the cavity resonance we derive a quality factor $Q=390 \pm 60$ corresponding to a cavity storage time of $\tau_{cav}=0.3 \pm 0.045 \ \rm{ps}$.

A versatile setup described in Ref. \cite{tijmen.2009.rsi} is used to Kerr-switch our microcavity. The setup is shown in Fig. \ref{yuce-f2}(a) and consists of two independently tunable optical parametric amplifiers (OPA, Light Conversion Topas pumped by a 1 kHz oscillator) that are the sources of the pump and probe beams. The pulse duration of both OPAs is $\tau_{P} = 140 \pm 10 \ \rm{fs}$.  The time delay $\Delta t$ between the pump and the probe pulse is set by a delay stage with a resolution of $15 \ \rm{fs}$. The reflected signal from the cavity is detected with a nitrogen cooled InGaAs line array detector spectrometer. The measured transient reflectivity contains information on the cavity resonance during the cavity storage time and it should thus not be confused with the instantaneous reflectivity at the delay $\Delta t$. The measured transient reflectivity is a result of the probe light that impinges at delay $\Delta t$ which circulates in the cavity during the storage time and then detected which is integrated due to the relatively slow response time of the detector \cite{tijmen.2009.rsi}.  

The cavity is switched with the electronic Kerr effect by judicious tuning of the pump and the probe frequencies relative to the semiconductor bandgap \cite{alex.2008.jap, harding.2009.josab}. The probe frequency ($\omega_{pr}= 7812 \ \rm{cm^{-1}}$) is set by the cavity resonance in the telecom range while the pump frequency is centered at $\omega_{pu}= 4165 \ \rm{cm^{-1}}$ $(\lambda_{pu}=2400 \rm{nm})$ to suppress nondegenerate two-photon absorption ($E_{pr}+E_{pu}\leq E_{gap}$) see Fig. \ref{yuce-f2}(b2). Furthermore, the energy of the pump photons is chosen to lie below half of the semiconductor band gap energy ($E_{pu}<\frac{1}{2}E_{gap}$), see Fig. \ref{yuce-f2}(b1), to avoid two pump-photon absorption. The excitation of free carriers is also suppressed by choosing a low probe pulse energy ($I_{pr}=0.18 \pm0.02 \rm{\ pJ / \mu m^2}$) while the average pump pulse energy is varied between $I_{pu}=13 \ \rm{and} \ 275 \pm 20 \rm{\ pJ / \mu m^2}$. The pulse energies are determined from the average laser power at the sample position and converted to peak power assuming a Gaussian pulse shape. The pulse energies are given per square micron since the switching of a cavity resonance has interesting prospects for miniature cavities with footprints in the micron range \cite{georgios.2010.prb, fushman.2007.apl}. The pump beam has a larger Gaussian focus ($\diameter _{pu}= 70 \ \rm{\mu m}$) than the probe beam ($\diameter _{pr}=30 \ \rm{\mu m}$) to ensure that only the central flat part of the pump focus is probed and that the probed region is spatially homogeneously pumped. The judicious selection of the pump-probe powers and frequencies enabled the instantaneous Kerr switching at elevated frequencies including the telecom band. 
 
\section{Pump pulse energy dependent ultimate-fast switching}

\begin{figure}
  \begin{center}
  \includegraphics[width=8.2 cm]{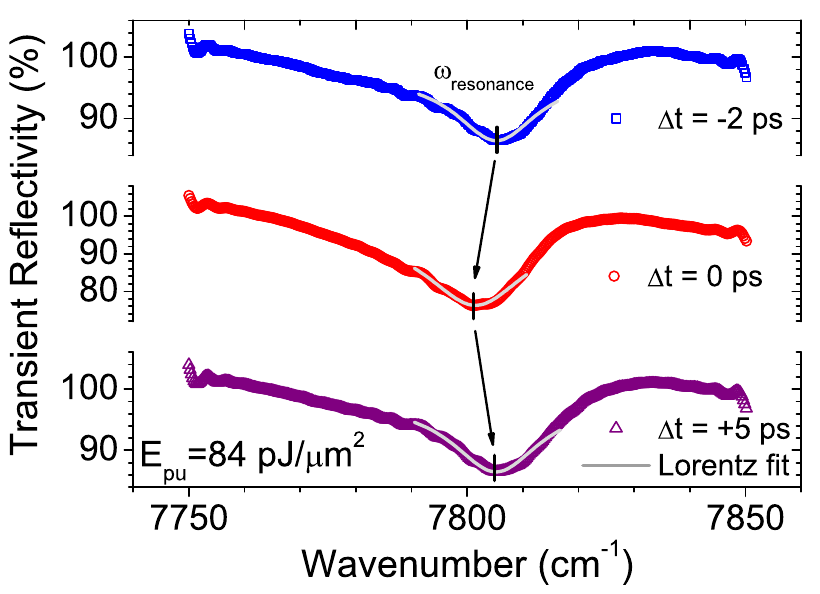}
  \caption{(color online) \emph{Transient reflectivity spectra for three different pump probe delays. The spectra are obtained at $84 \ pJ/\mu m^2$ pump pulse energy. The orange curves show the  fit to the cavity resonance from which the cavity resonance frequency ($\omega_{res}$) is determined as the minimum.}}
\label{yuce-f3}
  \end{center}
\end{figure}

Figure \ref{yuce-f3} shows the transient reflectivity spectra for three different pump-probe time delays $\Delta t$. At $\Delta t=-2 \ \rm{ps}$ the probe pulse arrives earlier than the pump pulse. Hence, the measured spectra shows the unswitched transient reflectivity of the cavity with a resonance at $\omega_{res}= 7805.6 \ \rm{cm^{-1}}$. At temporal overlap of pump and probe pulses ($\Delta t=0 \ \rm{ps}$) the cavity resonance frequency has red-shifted to $7800.7 \ \rm{cm^{-1}}$ indicating an increase of the refractive index. At positive delays ($\Delta t=+5 \ \rm{ps}$), where the pump pulse arrives earlier than the probe pulse, the cavity resonance is measured at $7805.2 \ \rm{cm^{-1}}$. Thus the resonance frequency at positive delays has returned to the same frequency as the unswitched resonance frequency at negative delays. The simultaneous observation of a red-shift of the cavity resonance only at pump-probe overlap and of the identical cavity resonances at positive and negative delays confirms that the cavity resonance is mainly switched by the electronic Kerr effect, and not by the free carriers.

\begin{figure}
  \begin{center}
  \includegraphics[width=8.2 cm]{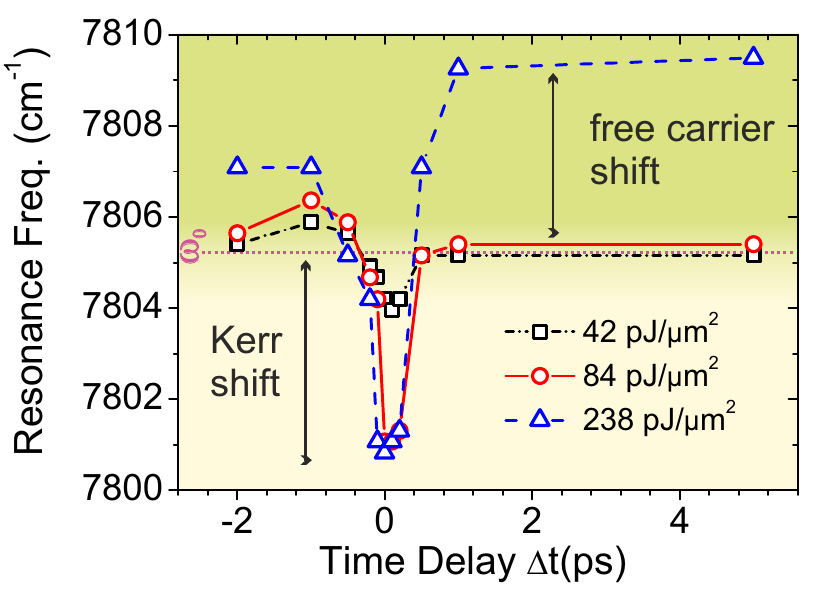}
  \caption{(color online) \emph{Resonance frequency versus time delay ($\Delta t$) between pump and probe at different pump pulse energies. The resonance frequency red-shifts due to the instantaneous electronic Kerr effect only at temporal overlap ($\Delta t=0 \ \pm 15 \ fs$) of pump-probe (shaded with bright color). The blue shift of the cavity resonance due to free carriers is observed when the pump pulse energy is increased (shaded with dark color). The dotted horizontal line shows the unswitched resonance frequency at $\omega_{0}=7805.6 \ \rm{cm^{-1}}$ }}
\label{yuce-f4}
  \end{center}
\end{figure}

Figure \ref{yuce-f4} shows the resonance frequency versus time delay at three different pump-pulse energies. Figure \ref{yuce-f4} is obtained from spectra similar to those shown in Fig. \ref{yuce-f3}. When the sample is pumped at $42 \pm 5 \ \rm{ pJ/\mu m^2}$ the resonance frequency red shifts by $1 \ \rm{cm^{-1}}$ at $\Delta t=0$. We observe the dynamic red-shift of the cavity resonance only at pump-probe coincidence within 300 fs, confirming the instantaneous switching of the cavity resonance frequency \citep{georgios.2011.apl}. The red-shift of the resonance frequency induced by the electronic Kerr effect increases to $4 \ \rm{cm^{-1}}$ when the cavity is pumped at $84 \pm 10 \ \rm{pJ/\mu m^2}$. At these power levels the cavity resonance frequency at positive time delays returns to the same value as the unswitched cavity resonance. Further increasing the pump pulse energy to $238 \pm \ 20 \ \rm{pJ/\mu m^2}$ results in an instantaneous shift of only $4.3 \ \rm{cm^{-1}}$ although the sample is pumped with a three times higher pulse energy. At high pump energies we also observe that the resonance frequency is blue shifted at positive time delays. At positive delays the refractive index decreases as a result of free carriers that remain excited for a much longer time (about $50 \ \rm{ps}$) as has been observed before \citep{harding.2007.apl, driel.2005.josab, wong.2009.apl}. The carriers are excited by two- and three-photon processes as depicted in Fig. \ref{yuce-f2} (b) and (c). Moreover, at high pump energies the cavity resonance frequency at $\Delta t<0 \ \rm{ps}$ is already blue shifted compared to the cold cavity resonance at low pump pulse energies, likely since the light is stored in the cavity up to  $\Delta t=-2 \ \rm{ps}$ which results in free carrier excitation by nondegenerate two- and three-photon absorption. We conclude from Fig. \ref{yuce-f4} that the reversible and ultrafast cavity switching as a result of the electronic Kerr effect occurs mainly at low pump pulse energies ($13-50 \pm 5 \rm{\ pJ / \mu m^2}$). In this regime, the switching speed is only limited by the cavity storage time and not by material relaxation properties. The switching of the cavity can be achieved within 300 fs, which is only limited by the storage time of light in the cavity and not by extrinsic material properties.

\begin{figure}
  \begin{center}
  \includegraphics[width=8.2 cm]{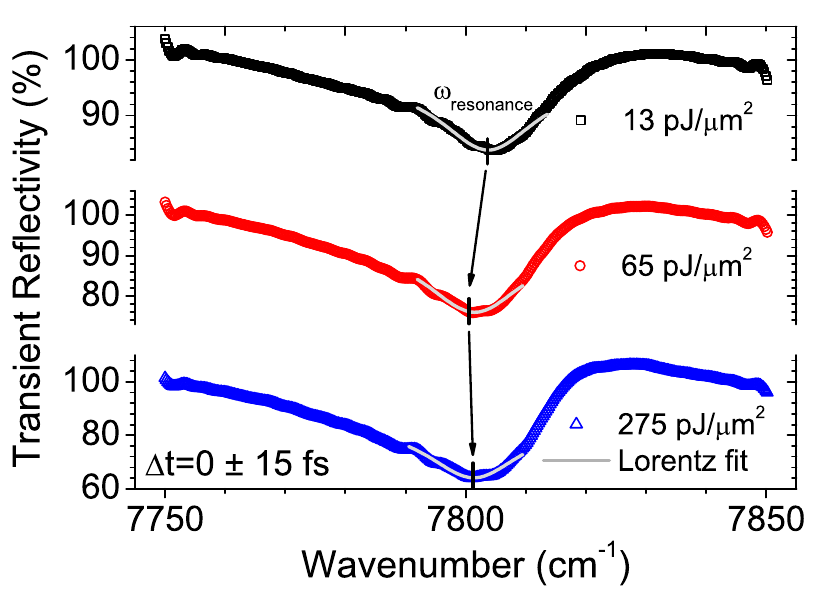}
  \caption{(color online) \emph{Transient reflectivity versus wavenumber for three different pump pulse energies. The spectra are obtained at pump-probe coincidence ($\Delta t=0 \pm 15 \ fs$). The orange curves show the fit to the cavity resonance from which the cavity resonance frequency ($\omega_{res}$) is determined as the minimum, indicated as ticks.}}
\label{yuce-f5}
  \end{center}
\end{figure}

Figure \ref{yuce-f5} shows the transient reflectivity spectra for three different pump-pulse energies at temporal ($\Delta t=0 \ \rm{ps}$) and spatial overlap of the pump and the probe beams. The cavity resonance frequency ($\omega_{res}= 7805.4 \ \rm{cm^{-1}}$), shifts to a lower frequency ($7804.2 \ \rm{cm^{-1}}$) if the sample is pumped at a low pump-pulse energy $13 \pm 1 \ \rm{ pJ / \mu m^2}$. When the pump-pulse energy is increased to $65 \pm 7 \ \rm{pJ / \mu m^2}$ we observe that the instantaneously switched resonance frequency further red-shifts to $7800.9 \ \rm{cm^{-1}}$. However, when the pump-pulse energy is increased to $275 \pm 20 \ \rm{pJ / \mu m^2}$ the resonance frequency shifts to only $7801.3 \ \rm{cm^{-1}}$ showing less shift than the previous step. We conclude that at high pump-pulse energies a competition takes place between the electronic Kerr effect that increases the refractive index and red-shifts the cavity resonance with the excited free carriers that decrease the refractive index and blue-shift the cavity resonance.

\begin{figure}
  \begin{center}
  \includegraphics[width=8.2 cm]{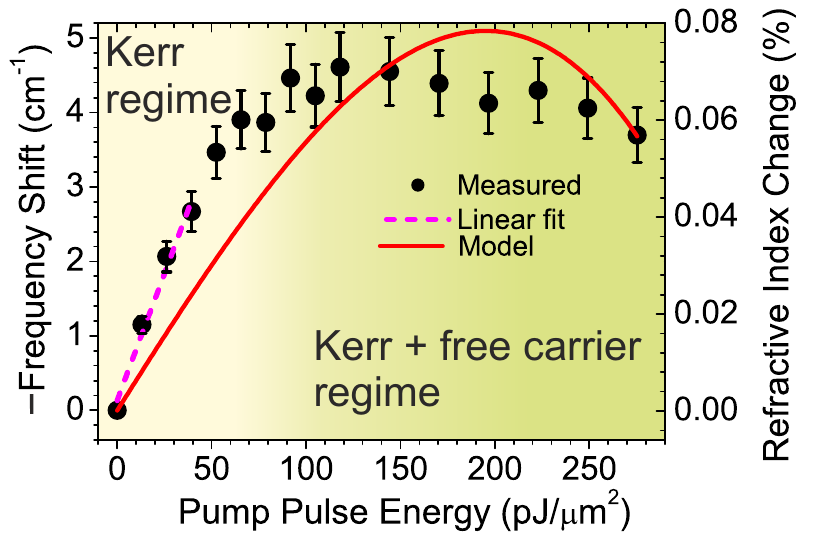}
  \caption{(color online) \emph{Instantaneous negative shift of the resonance frequency versus pump-pulse energy. Black circles show the measured results with a 10$\%$ error bar. At low pump-pulse energies (bright shaded region) we only observe electronic Kerr effect since the resonance frequency decreases linearly with the pump pulse energy (magenta dashed line). The competing blue shift of free carriers is observed beyond $70 \ pJ / \mu m^2$ and the region with both Kerr and free carrier excitation is darker shaded. The right ordinate shows the calculated refractive index change. The red solid curve indicates the modelled index change as a function of the pump-pulse energy for general nondegenerate pump-probe light beams.}}
\label{yuce-f6}
  \end{center}
\end{figure}

Figure \ref{yuce-f6} shows the shift of the instantaneously switched resonance frequency  versus pump-pulse energy. We observe that at low pump pulse energies the resonance frequency shift increases linearly with pump-pulse energy due to the positive refractive index  change of the electronic Kerr effect \citep{boyd.1992}. The linear increase of both the cavity resonance and the refractive index is physically reasonable since the nonlinear index change with the Kerr effect is a product of the nonlinear susceptibility with the pump field squared (see section modeling $\&$ Ref \cite{boyd.1992}). Beyond $50 \pm 5 \rm{\ pJ / \mu m^2}$ we observe a saturation and even a turn over of the resonance frequency shift versus pump-pulse energy. At high pump pulse energies, free carriers are excited that reduce the refractive index, opposite to the Kerr effect. Since the carriers with our settings of light frequencies can only be excited by two- and three-photon processes, the dependence of the refractive index (and hence resonance frequency) becomes nonlinear versus pump-pulse energy as is apparent from Fig. \ref{yuce-f6}. We see in Fig.~\ref{yuce-f6} that there is an apparent saturation in our experimental results. Our model does not show this saturation but a maximum. Since the pulse energies are limited in our experiments we cannot observe a possible decrease of the refractive index at high pulse energies. Moreover, in our model we do not consider the plasma screening effect and the Stark effect. From the linear slope at low pump energies where free carriers are negligible, we derive a \textit{nondegenerate} third-order nonlinear susceptibility $\chi^{(3)}=0.48\times10^{-11} \ \rm{esu}$ for GaAs at the strongly nondegenerate conditions $\omega_{pr}= 7812 \ \rm{cm^{-1}}$ and $\omega_{pu}= 4165 \ \rm{cm^{-1}}$ (dashed line in Fig. \ref{yuce-f6}). The value that we find for $\chi^{(3)}$ agrees within an order of magnitude with degenerate values reported at $\omega=9434 \ \rm{cm^{-1}}$ \cite{boyd.1992}. The qualitative agreement between these different measurements is  gratifying in view of the differences in the frequencies of the light sources~\cite{bristow.2007.apl}.

\section{Modelling}

In order to interpret the competition between the electronic Kerr effect and free carrier effects at elevated pump-pulse energies, we have developed a model of the optical properties of the semiconductor cavity. Here, we notably develop a model to include the nondegenerate three photon absorption. Since only the two-photon absorption coefficient has been reported previously for the nondegenerate case ~\cite{hutchings.1992.josab}, we chose to develop a model for nondegenerate case that can predict \textit{both} two- and three-photon absorption cross sections. In our experiment, as a result of the cavity field enhancement the effect of the probe intensity becomes comparable to the pump intensity. For this reason, we chose to develop a new model instead of using the existing degenerate three-photon absorption models~\cite{wherrett.1984.josab} to calculate the free carrier density as a function of both pump and probe intensities \textit{independently}. Our model describes the refractive index change of a cavity resulting from both the electronic Kerr effect and the free carriers in a nondegenerate pump-probe experiment. In order to model the refractive index change induced by the electronic Kerr effect we use the $\chi^{(3)}$ value determined from our experiments to limit the number of free parameters. For free carriers the relevant $\chi^{(3)}$ has been theoretically described using the independent particle approximation \cite{sipe.1194.prb, sipe.1995.prb}. The index change induced by free carriers that are excited by two (Fig. \ref{yuce-f2}(b)) and three (Fig. \ref{yuce-f2}(c)) photon absorption is calculated using the well-known Drude model \cite{drude.1900, drude2.1900}. We can safely neglect the contribution of free carriers generated by one-photon absorption since both the pump and the probe photons are much less energetic than the band-gap energy of GaAs. We combine both the electronic Kerr effect and the index change resulting from excited free carriers in Eq. \ref{1index}, which gives the refractive index change $\Delta n$ for a semiconductor optically switched in a general nondegenerate pump-probe experiment.  

 \begin{equation}
  {\Delta n} = \underbrace{{6 \pi \chi^{(3)} \over n_0}[\vert E_{cav}\vert^2+2 \vert E_{pu}\vert^2]}_{Kerr} - \underbrace{{q^2 \over 2n_0\epsilon_0 m_{opt}^* \omega_{pr}^2} [N_{eh}^{(2)}+N_{eh}^{(3)}]}_{free \ carriers},
 \label{1index}
 \end{equation}

Here, $\chi^{(3)}$ is the third order nonlinear susceptibility, $\epsilon_0$ the vacuum permittivity, $E_{cav}$ and $E_{pu}$ the electric field of the probe in the cavity and of the pump pulses, respectively, $q$ the electron charge, $\omega_{pr}$ and $\omega_{pu}$ the frequencies of the probe and the pump beams, respectively, $m_{opt}^*$ the optical effective mass of the free carriers, $N_{eh}^{(2)}$ and  $N_{eh}^{(3)}$ the free carrier densities generated by two-photon and three-photon absorption, respectively. Eq. \ref{1index} shows that the refractive index increases with the electronic Kerr effect and decreases with the increasing density of free carriers. In Eq. \ref{1index} the refractive index change induced by the electronic Kerr effect depends on the square of the electric fields of \textit{both} the pump and the probe light (the factor 2 for $E_{pu}$ will become clear at Eq. \ref{3polarization}). The electric fields in the instantaneous (Kerr) part of Eq. \ref{1index} can be written in terms of cycle-averaged intensity using the relation $I^*= \vert E \vert^2  n_0 c/ 2 \pi$. In general, when describing the intensity-dependent refractive index the effect of the probe is neglected due to its smaller intensity compared to the pump \cite{boyd.1992}. However, this is not necessarily the case for cavities due to the resonant field enhancement inside the cavity, which appears to be the case in our experiment. In general the cavity enhancement is given by ${I^{*}_{cav}/I^{*}_{pr}}=Q/2\pi\sqrt{R}$, where Q is the quality factor of the cavity and R the reflectivity of the cavity mirrors \cite{kavokin.2007}. The probe-pulse energy in our experiments is around $I^*_{pr}=0.18 \pm \ 0.02 \ \rm{pJ / \mu m^2}$ before entering the cavity.  Our cavity with Q=390 enhances the probe field by $Q/2\pi\sqrt{0.98}=63$ times to $I^*_{cav}=11.3 \pm \ 0.02 \ \rm{pJ / \mu m^2}$ so that it becomes non-negligible compared to the typical pump pulse energy ($I^*_{pu} \sim 10^{2}\ \rm{pJ / \mu m^2}$) in the cavity. The resonant enhancement of the probe pulses by the cavity becomes even more important for high quality factor cavities \cite{noda.2003.nature, vahala2.2003.nature, notomi.2010.nat.ph}, which might even bring the probe intensity beyond the level of the pump intensity. As a result, if the effect of probe light is neglected (see Eq. \ref{1index}) the usual pump intensity-dependent refractive index change will be incorrect especially for high quality factor cavities. 

In order to calculate the intensity-dependent refractive index for the general nondegenerate pump-probe case we start by writing the total optical field as $\tilde{E}(t)=E_{cav}(\omega_{pr})e^{-i \omega_{pr} t}+E_{pu}(\omega_{pu})e^{-i \omega_{pu} t}+c.c.$. The general form of the total polarization of a material is described up to third order by 
\begin{equation}
  {P^{Tot}(\omega)=\epsilon_0  \chi^{(1)} \tilde{E}(\omega)+\epsilon_0  \chi^{(2)} \tilde{E}^2(\omega)+ \epsilon_0 \chi^{(3)} \tilde{E}^3(\omega)}.
 \label{2polarization}
 \end{equation}
Due to the centrosymmetry of the GaAs $\chi^{(2)}=0$ the total polarization of the material reduces to  $P^{Tot}(\omega)=\epsilon_0 \chi^{(1)} \tilde{E}(\omega) + \epsilon_0  \chi^{(3)} \tilde{E}^3(\omega)$ \citep{yablonovitch.1972.prl}. Taking the third power of the total optical field for nondegenerate pump-probe light and inserting it into the total polarization leads to 
\begin{equation}
  P^{Tot}(\omega_{pr})=\epsilon_0 \chi^{(1)} E_{cav}e^{-i \omega_{pr} t} +3 \epsilon_0 \chi^{(3)}  E_{cav}^3e^{-i (\omega_{pr}+\omega_{pr} - \omega_{pr}) t}+6\epsilon_0 \chi^{(3)}  E_{cav} E_{pu}^2e^{-i( \omega_{pu} -\omega_{pu}+ \omega_{pr}) t},
 \label{3polarization}
 \end{equation}
which is the nonlinear polarization that influences the propagation of a beam of frequency $\omega_{pr}$. The two-fold degeneracy factor in front of $\vert E_{pu} \vert^2$ in Eq. \ref{1index} is due to the two-fold coefficient of the last term in Eq. \ref{3polarization} (see Appendix A for derivation). With this correction our model takes the cavity field enhancement into account and gives the appropriate solution for the intensity dependent refractive index for comparable intensities of nondegenerate pump-probe light. Having explained the Kerr term of Eq. \ref{1index} we calculate the free carrier term of Eq. \ref{1index} by calculating the free carrier densities as follows:

 \begin{eqnarray}
  N_{eh}^{(2)} & =  R_{ng}^{(2)}N_{atm}\tau_{int}= & {N_{atm}\tau_{int}8 \pi^{3}\vert \mu_{mg} \mu_{nm}\vert^2 \over \hbar^2 n_{0}^{2}c^2} 
\Bigg[  \overbrace{I_{cav}^2\rho_f(\omega_{ng}=2\omega_{pr})}^{Fig. \ref{yuce-f2}(b3)}  \nonumber \\
&& + \overbrace{I_{cav}I_{pu}\omega_{pr}\omega_{pu}\rho_f(\omega_{ng}=\omega_{pr}+\omega_{pu})\left(\frac{1}{\omega_{pr}^2}+\frac{2}{\omega_{pr} \omega_{pu}}+\frac{1}{\omega_{pu}^2} \right)}^{Fig. \ref{yuce-f2}(b2)} \Bigg],
 \label{4twophotondensity}
 \end{eqnarray}
and
\begin{eqnarray}
N_{eh}^{(3)} & = & R_{lg}^{(3)}N_{atm}\tau_{int}={N_{atm}\tau_{int}16 \pi^{4}\vert \mu_{mg} \mu_{nm} \mu_{ln}\vert^2\over \hbar^3 n_{0}^{3}c^3}\Bigg[
\overbrace{\frac{I_{pu}^3\rho_f(\omega_{lg}=3\omega_{pu})}{4\omega_{pu}}}^{Fig. \ref{yuce-f2}(c1)} +\overbrace{\frac{I_{cav}^3\rho_f(\omega_{lg}=3\omega_{pr})}{4\omega_{pr}}}^{Fig. \ref{yuce-f2}(c4)}  \nonumber \\
&& + \overbrace{I_{cav}I_{pu}^2\omega_{pr}\omega_{pu}^2\rho_f(\omega_{lg}=\omega_{pr}+2\omega_{pu}) \bigg(\frac{1}{4\omega_{pu}^4}+\frac{1}{\omega_{pu}^2 (\omega_{pr} +\omega_{pu})^2}+ \frac{1}{\omega_{pr}^2 (\omega_{pr}+\omega_{pu})^2} }^{Fig. \ref{yuce-f2}(c2)} \nonumber \\ 
&& + \overbrace{ \frac{1}{\omega_{pr} \omega_{pu}^2(\omega_{pr}+\omega_{pu})}+\frac{1}{\omega_{pu}^3(\omega_{pr}+\omega_{pu})}+\frac{2}{\omega_{pr}\omega_{pu}(\omega_{pr}+\omega_{pu})^2} \bigg) }^{Fig. \ref{yuce-f2}(c2)} \nonumber \\ 
&& + \overbrace{ I_{cav}^2I_{pu}\omega_{pr}^2\omega_{pu} \rho_f(\omega_{lg}=2\omega_{pr}+\omega_{pu}) \bigg(\frac{1}{4\omega_{pr}^4}+\frac{1}{\omega_{pu}^2 (\omega_{pr} +\omega_{pu})^2}  + \frac{1}{\omega_{pr}^2 (\omega_{pr}+\omega_{pu})^2} }^{Fig. \ref{yuce-f2}(c3)} \nonumber \\ 
&&+ \overbrace{ \frac{1}{\omega_{pr}^2 \omega_{pu}(\omega_{pr}+\omega_{pu})} +\frac{1}{\omega_{pr}^3(\omega_{pr}+\omega_{pu})}+\frac{2}{\omega_{pr}\omega_{pu}(\omega_{pr}+\omega_{pu})^2}\bigg) }^{Fig. \ref{yuce-f2}(c3)} \Bigg].
 \label{5threephotondensity}
 \end{eqnarray}
 
The two- and three-photon free carrier densities ($N_{eh}^{(2)}$ and $N_{eh}^{(3)}$) in Eqs. \ref{4twophotondensity} and \ref{5threephotondensity} are calculated by multiplying the excitation rate $R_{eh}$ with the number of atoms $N_{atm}$ in the unit volume and the interaction time $\tau_{int}$ (see Appendix B and C for derivations of general case). The relatively slow response time of the free carriers and the accumulation nature of the free carrier excitation will integrate in time and this will mask the ultrafast dynamics. Therefore, the intensities in Eqs. \ref{4twophotondensity} and \ref{5threephotondensity} are defined as $I=\int \vert E \vert^2  n_0 c/ 2 \pi dt$, which differs from the instantaneous intensity term ($I^*$) in the Kerr term of Eq. \ref{1index}. The limits of the time integral are given by the duration time of the excitation process, which is much longer than the excitation time of the carriers. The pump interaction time is given by the pulse duration $\tau_{P}$ whereas the probe interaction time is given by $\tau_{cav}$ due to the cavity which is in resonance with the probe light only. In Eqs. \ref{4twophotondensity} and \ref{5threephotondensity}, $\rho_f(\omega_{ng})$ and $\rho_f(\omega_{lg})$ are the density of final states and $\mu_{mg}, \mu_{nm}, \mu_{ln}$ are the dipole transition moments associated with the resonance schemes depicted in Fig. \ref{yuce-f2}(b) and (c). The parameters used in our model are listed in table~\ref{table1}.

\begin{table}[h]
\caption{Parameters used in our model.}  
\centering
\footnotetext{$^\dagger$ Set by the experimental conditions.}
\begin{tabular}{p{3cm} r@{.}l l r}
\hline\hline                      
Parameter & & Value& Unit & Source \\ [1.0ex]   
\hline        
$N_{atm}$ & 4 & $42\times 10^{22}$ & $\rm{atoms/cm^3}$ & \cite{deepa.2010} \\
$\mu^2$ & 6&$25\times10^{-43}$ & $\rm{Jcm^{3}}$ & \citep{boyd.1992} \\
$\Gamma_{n,l}$ & $6$&$28\times10^{13}$ & rad/s & \citep{boyd.1992} \\
$\tau_{int}$ & 150 &$0\times10^{-15}$ & s & $^\dagger$ \\
$\omega_{pr}$ & 7805 &$79$ & $\rm{cm^{-1}}$ & $^\dagger$ \\
 & 1 &$47\times10^{15}$ & Hz & $^\dagger$ \\
$I^*_{pr}$ & 0 &$18$ & $\rm{pJ / \mu m^2}$ & $^\dagger$ \\
 & 0 &$2$ & $\rm{GW /cm^2}$ & $^\dagger$ \\
$I^*_{cav}$ & 11 &$3$ & $\rm{pJ / \mu m^2}$ & $^\dagger$ \\
 & 12 &$0$ & $\rm{GW /cm^{2}}$ & $^\dagger$ \\
$\omega_{pu}$ & 4166 &$67$ & $\rm{cm^{-1}}$ & $^\dagger$ \\
 & 7 &$85\times10^{15}$ & $\rm{Hz}$ & $^\dagger$ \\
$I^*_{pu}$ & 1 &$00\times10^{2}$ & $\rm{pJ / \mu m^2}$ & $^\dagger$\\
 & 93 &$7$ & $\rm{GW /cm^{2}}$ & $^\dagger$ \\[1ex]
\hline
\end{tabular}
\label{table1}
\end{table}

We have derived the two-photon absorption rate $R^{(2)}_{ng}$ using the perturbation solution to the Schr\"odinger's equation for nondegenerate applied optical fields, see Appendix B. We calculate the two-photon absorption cross section $\sigma_{ng}^{(2)}$ by taking into account that the energy of two pump photons is less than the electronic bandgap energy of GaAs [$2 \times E_{pu}(0.51 \ \rm{eV}) < E_{gap}(1.43 \ \rm{eV})$, Fig. \ref{yuce-f2}(b1)] so that we can safely neglect the excitation of free carriers by two-pump photons. We consider absorption of two probe photons (Fig. \ref{yuce-f2}(b3)) and the nondegenerate two-photon absorption (Fig. \ref{yuce-f2}(b2)) when we calculate the two-photon absorption cross section. Under the circumstances listed in Table~\ref{table1} we calculate the two-photon absorption cross section to be equal to $\sigma_{ng}^{(2)}(\omega_{pr},\omega_{pu})=1.14 \times 10^{-50} \ \rm{cm^4 s / photons^2}$. The two-photon absorption coefficient ($\beta^{(2)}$) can be calculated from $\beta^{(2)}=4\sigma_{ng}^{(2)}N_{atm}/\hbar\omega$~\citep{nathan.1985.josab}. We obtain the two-photon absorption coefficient to be $\beta^{(2)}=0.013\ \rm{cm/MW}$. The values for $\sigma_{ng}^{(2)}$ and $\beta^{(2)}$ agree with the earlier estimated and measured values \cite{xu.1997, nathan.1985.josab}. In our experiment the sum of the energies of the pump and the probe photons are chosen to suppress two-photon absorption $[E_{pr}(0.95 \ \rm{eV})+E_{pu}(0.51 \ \rm{eV})\simeq E_{gap}(1.43 \ \rm{eV})]$ which affects the two-photon absorption cross section. We adjust the two-photon cross section to $\sigma_{ng}^{(2)*}(\omega_{pr},\omega_{pu})=8.6 \times 10^{-52} \ \rm{cm^4 s / photons^2}$ to obtain a good match of our model with our experimental data. With the value we use for $\sigma_{ng}^{(2)*}$ we observe that the refractive index increases as in our experiment with the applied pump pulse energy. We use the value that we calculate for $\sigma_{ng}^{(2)*}(\omega_{pr},\omega_{pu})$ in our model (solid curve in Fig. \ref{yuce-f6}) to calculate the two-photon generated free carrier density. As in the experiments, we observe that the linear increase of the index change due to the electronic Kerr effect competes with the excited free carriers whose density increases linearly with pump-pulse energy (since $E_{pu}<\frac{1}{2}E_{gap}$). The nondegenerate two-photon coefficient can also be estimated for GaAs using the model described by Hutchings et al~\cite{hutchings.1992.josab}. However, we calculate the nondegenerate two-photon absorption coefficient using our model since it was a necessary step to calculate nondegenerate three-photon absorption coefficient. 

We have derived the density of free carriers (Eq. \ref{5threephotondensity}) generated by three-photon absorption $N_{eh}^{(3)}$. We calculate the three-photon absorption rate $R^{(3)}_{lg}$ using the perturbation solution to the Schr\"odinger's equation for nondegenerate optical fields (see Appendix C). The three-photon absorption cross section $\sigma_{lg}^{(3)}$ is calculated by considering all excitation schemes shown in Fig. \ref{yuce-f2}(c), since all the permutations of pump and probe exceed the bandgap of GaAs. We find the three-photon absorption cross section to be $\sigma_{lg}^{(3)}(\omega_{pr},\omega_{pu})=3.9 \times 10^{-84} \ \rm{cm^6 s^2 / photons^3} $. The three-photon absorption coefficient ($\gamma^{(3)}$) can then be calculated from $\gamma^{(3)}=6\sigma_{lg}^{(3)}N_{atm}/(\hbar\omega)^2$~\citep{nathan.1985.josab}. We obtain the three-photon absorption coefficient to be $\gamma^{(3)}= 0.45\times 10^{-4} \ \rm{cm^3/GW^2}$. The values we calculated for $\sigma_{lg}^{(3)}$ and $\gamma^{(3)}$ agree within two \cite{xu.1997} and one \cite{nathan.1985.josab} order of magnitude with the reported values. We consider this a very good agreement in the view of the difference in frequencies, material and the difference due to degenerate and nondegenerate conditions. The value that we experimentally determine for $\gamma^{(3)}$ is in good agreement with the measured values~\cite{he.2005.oe}. We set the three-photon absorption cross section to $\sigma_{lg}^{(3)*}(\omega_{pr},\omega_{pu})=4.7 \times 10^{-83} \ \rm{cm^6 s^2 / photons^3}$ to obtain a good match of our model with our experimental data. With the value we use for $\sigma_{lg}^{(3)*}(\omega_{pr},\omega_{pu})$ the refractive index change becomes nonlinear with the applied pump-pulse energy and the refractive index starts to decreases within the energy regime shown in Fig. \ref{yuce-f6}. The model also clearly shows the desired linear increase of the index in the Kerr regime (at low pulse energies) and the appearance of the nonlinear decrease of the free-carrier index that starts to compete at higher pulse energies. We attribute the difference between the model and the experiment at low pump pulse energies to our choice of not using $\chi^{(3)}$ as a free parameter while we consider the free carrier excitation even at low pump-pulse energies in our model. In our analysis we do not calculate the density of states for each permutation of pump-probe frequencies although our model can describe the frequency dependency. Instead we use the approximation that $\rho_f(\omega) \approx (2 \pi \Gamma_{n,l})^{-1}$ \citep{boyd.1992} where $\Gamma_{n,l}$ (see Table~\ref{table1}) is the width of level \textit{n, l}.  We list all the coefficients that we calculate using our model and the coefficients that we deduce from our experiment in Table~\ref{table2}. We conclude here that there is an optimum power for instantaneous Kerr switching of a cavity, namely at the onset of the carrier effects this value optimizes.

\begin{table}[h]
\caption{Coefficients calculated and determined from measurements. \\ The values are calculated and measured for the experimental parameters listed in table~\ref{table1}.}  
\centering  
\begin{tabular}{l c c rp{4cm}}  
\hline\hline                       
Parameter & Measurement & Calculated & Unit \\ [1.0ex]   
\hline              
\raisebox{-1.5ex}{ $\chi^{(3)}$} & $6.72\times10^{-20}$ &  & $\rm{m^2/V^2}$  \\
 & \raisebox{1.5ex}{$0.48\times10^{-11}$} &  & \raisebox{1.5ex}{esu}  \\
$\sigma_{ng}^{(2)}$ & $8.59 \times 10^{-52}$ & $1.14 \times 10^{-50}$ & $\rm{cm^4 s / photons^2}$ \\

$N_{eh}^{(2)}$ & $3.78 \times 10^{17}$ & $5.04 \times 10^{18}$ & $\rm{1/cm^3}$ \\

$\beta^{(2)}$ & $0.10\times 10^{-2}$ & $0.13\times 10^{-1}$ & $\rm{cm/MW}$ \\

$\sigma_{lg}^{(3)}$ & $4.71 \times 10^{-83}$ & $3.93 \times 10^{-84}$ & $\rm{cm^6 s^2 / photons^3}$ \\

$N_{eh}^{(3)}$ & $5.92 \times 10^{16}$ & $4.93 \times 10^{15}$ & $\rm{1/cm^3}$ \\

$\gamma^{(3)}$ & $0.53\times 10^{-3}$ & $0.45\times 10^{-4}$ & $\rm{cm^3/GW^2}$ \\
\hline
\end{tabular}
\label{table2}
\end{table}

\section{Quality factor dependent ultimate-fast switching}

\begin{figure}
  \begin{center}
  \includegraphics[width=8.2 cm]{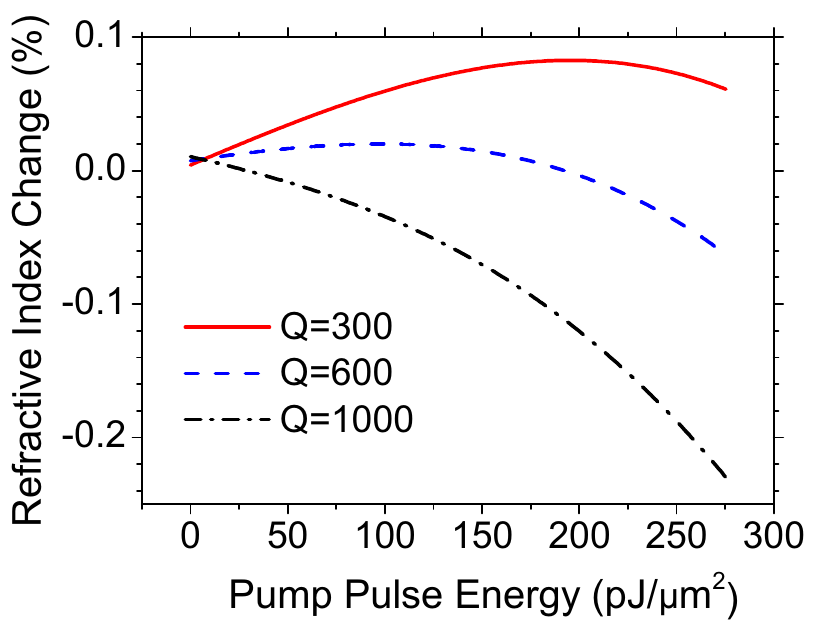}
  \caption{(color online) \emph{Refractive index change versus pump-pulse energy calculated for three different quality factors. The positive index change due to the electronic Kerr effect is more pronounced with low quality factor cavities with fast dynamics.  We assumed cavities with resonance $\omega_{res}=7812 \ \rm{cm^{-1}}$ pumped at $\omega_{pu}=4165 \ \rm{cm^{-1}}$ as in our experiment. The pump pulse duration is taken as $\tau_{P} = 140 \pm 10 \ \rm{fs}$, whereas the probe pulse duration is set by $\tau_{cav}$.}}
\label{yuce-f7}
  \end{center}
\end{figure}

As opposed to the derivation of three photon absorption coefficient ($\sigma_{lg}^{(3)}$) that restricts itself to the simplified case of degenerate optical fields \cite{wherrett.1984.josab, boyd.1992}, we have here derived the two- and three-photon absorption cross sections for the general case of nondegenerate optical fields. Since only the two-photon absorption coefficient has been reported previously for the nondegenerate case~\cite{hutchings.1992.josab} we chose to derive a model for nondegenerate case that can predict \textit{both} two- and three-photon absorption cross sections. Our approach holds the additional advantage of calculating the free carrier density as a function of both pump and probe intensities \textit{independently}. This feature allows us to extend refractive index changes for switched cavities with different quality factors Q. We assumed cavities with resonance $\omega_{res}=7812 \ \rm{cm^{-1}}$ pumped at $\omega_{pu}=4165 \ \rm{cm^{-1}}$ as in our experiment. The pump pulse duration is taken as $\tau_{P} = 140 \pm 10 \ \rm{fs}$, whereas the probe pulse duration is set by $\tau_{cav}$ since only the probe pulse is in resonance with the cavity and $\tau_{cav}$ is inversely proportional to the quality factor Q.  Figure \ref{yuce-f7} shows that the observed refractive index increase from the Kerr effect can be revealed with low quality factor (Q=300) cavity up to $200 \ \rm{ pJ / \mu m^2}$ pump pulse energy, similar to our experiment. For increasing quality factors there is only a small increase in the refractive index due to the Kerr effect before the free carriers decrease the index (Q=600) or even only a decreasing refractive index with increasing pump pulse energy (Q=1000). The less apparent Kerr effect with the increasing quality factor is caused by the decreasing temporal overlap of pump and probe as the probe pulse becomes much longer than the pump pulse ($\tau_{cav}>\tau_{P}$) \citep{georgios.2011.apl}. In fact, for high quality factor cavities during a longer fraction of the probe pulse there is no pump light, as a result no Kerr switching occurs for this time duration. As a consequence, high quality factor cavities invite Kerr switching with long pump pulses, but this defies the purpose of ultrafast optical switching. Interestingly, however, there is (Fig. \ref{yuce-f7}) also already Kerr-induced refractive index increase for zero pump-pulse energy. This effect is the result of the cavity enhanced probe light that already induces a Kerr-shift. With increasing quality factor, the shift increases because of the increased probe-enhancement in the cavity. However, due to the competing free carriers generated via degenerate two- and three- probe photon absorption the Kerr induced positive shift does not scale linearly with the quality factor. We therefore conclude that cavities with shorter storage times can reduce the free carrier excitation which enables instantaneous switching of semiconductor cavities at the telecom range.  

\section{Conclusion}
We demonstrate switching of a semiconductor microcavity within 300 fs at telecom wavelengths using the electronic Kerr effect as a function of pump-pulse energy. We manage to measure the nondegenerate third order susceptibility ($\chi^{(3)}$) of GaAs using pump-probe experiment. We show that the refractive index change induced by the electronic Kerr effect can be increased to a certain extent that is limited by the increasing density of excited free carriers. We show that the judicious tuning of the frequency of the driving fields relative to the band gap of the semiconductor decreases the number of free carriers and thereby increases the positive shift of the resonance frequency resulting from the electronic Kerr effect. Our model quantitatively describes the frequency and the intensity dependence of nondegenerate switching with pump-probe experiment. The realization and the understanding of the competition between the electronic Kerr effect and the free carriers reveals the set of parameters using which the instantaneous electronic Kerr effect can be utilized as the ultimate-fast way of all-optical switching. The refractive index change ($0.1 \%$) induced by the electronic Kerr effect will result in a larger resonance frequency shift in comparison to the cavity linewidth with high-Q cavities. However, we find that due to the larger field enhancements in high-Q cavities the Kerr effect will be hindered by the free carriers. On the other hand, if the incident probe pulse energy is further reduced then this competition in high-Q cavities can be directed in favour of the electronic Kerr effect. Reducing the required pulse energy of the pump pulses will be achieved if the pump pulses are also resonantly enhanced by the cavity. Moreover, we note that the pump photons are not absorbed in the electronic Kerr effect, hence the pump pulses do not heat up the sample and thus they can be recycled to switch the cavity resonance again.

\section{Appendix A: Intensity dependent refractive index}

In nonlinear optics the intensity dependent refractive index for pump-probe experiments is generally described under the assumption of weak probe field \cite{boyd.1992}. In experiments involving cavities however, as described in this paper, pump and probe fields may be of the same order due to the resonant enhancement of the probe field. In order to reveal the consequences of this, we derive in this appendix the intensity-dependent refractive index involving nondegenerate pump and probe fields.

The nonlinear refractive index can be described for large interaction times \citep{boyd.1992}
\begin{equation}
n(\omega_{pr})=n_0(\omega_{pr}) + n_{2}(\omega_{pr}:\omega_{pr},\omega_{pu})\langle \tilde{E}^{2}\rangle,
\label{6index}
\end{equation}
where $n_0$ is the weak field refractive index, $n_2$ the second-order index of refraction and $\langle \tilde{E}\rangle$ the time average of the electric field. The arguments of $n_2$ express that the result at frequency $\omega_{pr}$ depends on both $\omega_{pr}$ and $\omega_{pu}$, as we will see below.  In order to calculate the intensity dependent refractive index for nondegenerate pump-probe light we start with an optical field of the form 
\begin{equation}
\tilde{E}(t)=E_{pr}(\omega_{pr})e^{-i \omega_{pr} t}+E_{pu}(\omega_{pu})e^{-i \omega_{pu} t}+c.c.,
\label{7field}
\end{equation}
so that 
\begin{equation}
\langle \tilde{E}^2(\omega_{pr},\omega_{pu})\rangle = 2 E_{pr}(\omega_{pr})E^*_{pr}(\omega_{pr})+2 E_{pu}(\omega_{pu})E^*_{pu}(\omega_{pu})=2(\vert E_{pr} \vert^2 + \vert E_{pu} \vert^2).
\label{8averagefield}
\end{equation}
By inserting Eq. \ref{8averagefield} into Eq. \ref{6index} we rewrite the nonlinear index in terms of the pump and probe fields
\begin{equation}
n(\omega_{pr})=n_0(\omega_{pr}) + 2n_{2}(\omega_{pr}:\omega_{pr},\omega_{pu})(\vert E_{pr} \vert^2 + \vert E_{pu} \vert^2).
\label{9index}
\end{equation}

The general form of the total polarization of a material is described up to third order by \citep{boyd.1992}
\begin{equation}
  {P^{Tot}(\omega)=\epsilon_0  \chi^{(1)} \tilde{E} (\omega)+\epsilon_0  \chi^{(2)} \tilde{E}^2(\omega)+ \epsilon_0 \chi^{(3)} \tilde{E}^3(\omega)}.
 \label{10polarization}
 \end{equation}
Due to the centrosymmetry of GaAs the total polarization reduces to  $P^{Tot}(\omega)=\epsilon_0 \chi^{(1)} \tilde{E}(\omega) + \epsilon_0  \chi^{(3)} \tilde{E}^3(\omega)$ \cite{yablonovitch.1972.prl}. Taking the total optical field to the third power  (Eq. \ref{7field}) and inserting it into the total polarization leads to 
\begin{eqnarray}
P^{Tot}(\omega_{pr})&=&\epsilon_0 \chi^{(1)} E_{pr}e^{-i \omega_{pr} t} +3 \epsilon_0 \chi^{(3)}  E_{pr}^3e^{-i (\omega_{pr}+\omega_{pr} - \omega_{pr}) t}+6\epsilon_0 \chi^{(3)}  E_{pr} E_{pu}^2e^{-i( \omega_{pu} -\omega_{pu}+ \omega_{pr}) t} \nonumber \\
& = & \epsilon_0 E_{pr}e^{-i \omega_{pr} t} \underbrace{(\chi^{(1)} +3 \chi^{(3)} E_{pr}^2 +6 \chi^{(3)} E_{pu}^2)}_{\chi^{eff}},  
 \label{11polarization}
 \end{eqnarray} 
which is the nonlinear polarization that influences the propagation of a beam of frequency $\omega_{pr}$. We introduce an effective nonlinear susceptibility in Eq.  \ref{11polarization} given by
\begin{equation}
\chi^{eff}=\chi^{(1)} +3 \chi^{(3)} \vert E_{pr}\vert^2 +6 \chi^{(3)} \vert E_{pu}\vert^2.
\label{12effectivetensor}
\end{equation}
We note that it is generally true that \citep{boyd.1992}
\begin{equation}
n^2(\omega_{pr})=1+4 \pi \chi^{eff},
\label{13index}
\end{equation}
and by inserting Eq. \ref{9index} and \ref{12effectivetensor} into Eq. \ref{13index} we get
\begin{eqnarray}
&& n_{0}(\omega_{pr})^{2}+4n_2(\omega_{pr}:\omega_{pr},\omega_{pu})n_0(\omega_{pr})(\vert E_{pr} \vert^2 + \vert E_{pu} \vert^2) + \\ \nonumber 
&& 4n_2(\omega_{pr}:\omega_{pr},\omega_{pu})^2(\vert E_{pr} \vert^2 + \vert E_{pu} \vert^2 )^2 \\ \nonumber
&& = 1+4\pi\chi^{(1)} +12 \pi \chi^{(3)} \vert E_{pr} \vert^2 +24 \pi \chi^{(3)} \vert E_{pu} \vert^2.
\label{14indexsuscep}
\end{eqnarray}
Making the reasonable assumption that $n_2<<n_0$ and by equating the terms of the same order on each side of Eq. \ref{14indexsuscep} we find the relation between the linear and nonlinear refractive indices and the relevant susceptibilities as follows:
\begin{equation}
n_{0}(\omega_{pr})^{2}=1+4 \pi \chi^{(1)},
\label{15index}
\end{equation}
\begin{equation}
n_{2}(\omega_{pr}:\omega_{pr},\omega_{pu})= \frac{3 \pi \chi^{(3)}}{ n_0(\omega_{pr})}\frac{(\vert E_{pr} \vert^2 +2 \vert E_{pu} \vert^2)}{(\vert E_{pr} \vert^2 + \vert E_{pu} \vert^2)}.
\label{16index}
\end{equation}
Inserting Eq. \ref{16index} into Eq. \ref{6index} results in the nonlinear index
\begin{equation}
n(\omega_{pr})=n_0(\omega_{pr})+ \frac{6 \pi \chi^{(3)}}{ n_0(\omega_{pr})}(\vert E_{pr} \vert^2 +2 \vert E_{pu} \vert^2).
\label{17indexfinal}
\end{equation}
Equation \ref{17indexfinal} shows that the refractive index change induced by the electronic Kerr effect depends on the square of the electric fields of \textit{both} the pump and the probe light. For low values of $E_{pr}$, we get the usual expression for two degenerate beam case given in \citep{boyd.1992}. In the text we use $E_{cav}$ instead of $E_{pr}$ since only the probe pulse is in resonance with the cavity, which modifies the probe pulse duration and the intensity of the probe pulse. However, in the appendix the equations are derived for a general case where the probe pulse is non-resonant.

\section{Appendix B: Two-photon absorption cross section}

Following the derivation of refractive index change induced by the electronic Kerr effect we derive the two-photon absorption rate $R^{(2)}_{eh}$ using a perturbation approach to solve Schr\"odinger's equation for nondegenerate applied optical fields. We start with a two-level system to calculate the two-photon absorption rate and later we introduce density of states in order to mimic a semiconductor. In our derivation we choose to explicitly write out all terms instead of using the permutation operator for the probability amplitude as in \citep{boyd.1992}. In this way we can calculate the absorption rate as a function of both pump and probe intensities as an extension beyond the textbook \citep{boyd.1992}.

We start by writing the time-dependent Schr\"odinger equation in the presence of a time-dependent interaction potential $\tilde{V}(t)$. We then use the standard perturbation analysis as described in Ref. \cite{boyd.1992} to get:
\begin{equation}
\frac{da_m^{(N)}}{dt}=(i\hbar)^{-1}\sum_{l}a_l^{N-1}\tilde{V}_{lm}e^{-i\omega_{lm}t},
\label{18schrodinger}
\end{equation}
where $a_m^{(N)}$ is the probability amplitude of sate m with N interaction order and  $\tilde{V}_{lm}$ are the matrix elements of interaction Hamiltonian $\hat{V}$. We first calculate the linear absorption term, hence we set N=1. We assume that in the absence of any applied electric field the atoms are in the ground state \textit{g} (see Fig. \ref{yuce-f2} for the energy levels) so that $a_g^{0}(t)=1$ and $a_m^{0}(t)=0$ for $m \neq g$ at all times \textit{t} \citep{boyd.1992}. We then write $\tilde{V}_{mg}$ as:
\begin{equation}
\tilde{V}_{mg}=-\mu_{mg}[E_{pr}(\omega_{pr})e^{-i \omega_{pr} t}+E_{pu}(\omega_{pu})e^{-i \omega_{pu} t}+E^*_{pr}(\omega_{pr})e^{i \omega_{pr} t}+E^*_{pu}(\omega_{pu})e^{i \omega_{pu} t}],
\label{19potential1}
\end{equation} 
where $\mu_{mg}$ is the transition dipole moment between states m and g. Inserting Eq. \ref{19potential1} into Eq. \ref{18schrodinger} gives
\begin{eqnarray}
\frac{da_m^{(1)}}{dt}&=&-(i\hbar)^{-1}\mu_{mg}[E_{pr}(\omega_{pr})e^{i( \omega_{mg} -\omega_{pr})t}+E_{pu}(\omega_{pu})e^{i (\omega_{mg}-\omega_{pu})t} \nonumber \\ && +E^*_{pr}(\omega_{pr})e^{i (\omega_{pr}+\omega_{mg}) t}+E^*_{pu}(\omega_{pu})e^{i (\omega_{pu}+\omega_{mg}) t}].
\label{20schlevel1}
\end{eqnarray}
We drop the terms with $\omega_{pr}+\omega_{mg} \ \rm{and} \ \omega_{pu}+\omega_{mg}$ since they describe the process of stimulated emission. The neglect of the second terms is known as the rotating wave approximation. To get the probability amplitude for linear absorption we integrate Eq. \ref{20schlevel1}
\begin{eqnarray}
a_m^{(1)}(t)&=&-(i\hbar)^{-1}\mu_{mg}\int_0^t dt'[E_{pr}(\omega_{pr})e^{i( \omega_{mg} -\omega_{pr})t'}+E_{pu}(\omega_{pu})e^{i (\omega_{mg}-\omega_{pu})t'}] \nonumber \\
&=&\frac{\mu_{mg}E_{pr}}{\hbar (\omega_{mg}-\omega_{pr})}[e^{i( \omega_{mg} -\omega_{pr})t}-1]+\frac{\mu_{mg}E_{pu}}{\hbar (\omega_{mg}-\omega_{pu})}[e^{i( \omega_{mg} -\omega_{pu})t}-1].
\label{21problevel1}
\end{eqnarray}

In order to get the probability amplitude $a_n^{2}(t)$ for two-photon absorption (see Fig. \ref{yuce-f2}(b) for the energy levels) we describe $\tilde{V}_{nm}$ as:
\begin{equation}
\tilde{V}_{nm}=-\mu_{nm}[E_{pr}(\omega_{pr})e^{-i \omega_{pr} t}+E_{pu}(\omega_{pu})e^{-i \omega_{pu} t}+E^*_{pr}(\omega_{pr})e^{i \omega_{pr} t}+E^*_{pu}(\omega_{pu})e^{i \omega_{pu} t}],
\label{22potential2}
\end{equation}
We use Eqs. \ref{21problevel1} and \ref{22potential2} in Eq.\ref{18schrodinger} to get
\begin{eqnarray}
\frac{da_n^{(2)}}{dt}&=&-(i\hbar)^{-1} \sum_m a_m^{(1)} \times \tilde{V}_{nm} e^{-i\omega_{mn}t}.
\label{23schlevel2}
\end{eqnarray}
In Eq. \ref{23schlevel2} we again use the rotating wave approximation and we omit the complex terms which describe the stimulated emission as has been shown in Eq. \ref{20schlevel1}. Furthermore, in Eq. \ref{23schlevel2} we assume that single level \textit{m} dominates the sum so that the sum disappears. Thus we get:
\begin{eqnarray}
\frac{da_n^{(2)}}{dt}=-\frac{\mu_{mg} \mu_{nm}}{i\hbar^2} & \Bigg[ & \frac{E_{pr}^2}{(\omega_{mg}-\omega_{pr})}[e^{i( \omega_{mg}+\omega_{nm} -2\omega_{pr})t}-e^{i( \omega_{nm}-\omega_{pr})t}] \nonumber \\
&+& \frac{E_{pr} E_{pu}}{(\omega_{mg}- \omega_{pu})}[e^{i( \omega_{mg}+\omega_{nm} -\omega_{pr} - \omega_{pu})t}-e^{i( \omega_{nm}-\omega_{pr})t}] \nonumber \\
&+& \frac{E_{pr} E_{pu}}{(\omega_{mg}- \omega_{pr})}[e^{i( \omega_{mg}+\omega_{nm} -\omega_{pr} - \omega_{pu})t}-e^{i( \omega_{nm}-\omega_{pu})t}] \nonumber \\
&+& \frac{E_{pu}^2}{(\omega_{mg}-\omega_{pu})}[e^{i( \omega_{mg}+\omega_{nm} -2\omega_{pu})t}-e^{i( \omega_{nm}-\omega_{pu})t}]\Bigg].
\label{24schlevel2}
\end{eqnarray}
In Eq. \ref{24schlevel2} we use the identity $(\omega_{ng}=\omega_{nm}+\omega_{mg})$ and we assume that the one-photon transition is highly non-resonant so that  $(\omega_{mg}-\omega_{pr} \simeq \omega_{pr})$ and $(\omega_{mg}-\omega_{pu}\simeq \omega_{pu})$. The terms with $(\omega_{nm}-\omega_{pr}) \ \rm{and} \ (\omega_{nm}-\omega_{pu})$ give the transient response of the process so that they can be dropped in the consideration of Eq. \ref{24schlevel2} \citep{boyd.1992}. Finally, we integrate Eq. \ref{24schlevel2} up to time \textit{t} and then multiply by $t/t$ to make the denominators look similar to the exponents:
\begin{eqnarray}
a_n^{(2)}(t)=\frac{t\mu_{mg} \mu_{nm}}{\hbar^2} & \Bigg[ & \frac{E_{pr}^2 [e^{i( \omega_{ng}-2\omega_{pr})t}-1]}{\omega_{pr}(\omega_{ng}-2\omega_{pr})t} + \frac{E_{pu}^2 [e^{i( \omega_{ng}-2\omega_{pu})t}-1]}{\omega_{pu}(\omega_{ng}-2\omega_{pu})t}
\nonumber \\
&+& \left( \frac{E_{pr} E_{pu}}{\omega_{pu}}+\frac{E_{pr} E_{pu}}{\omega_{pr}} \right)  \frac{ e^{i( \omega_{ng}-\omega_{pr} -\omega_{pu})t}-1  }{( \omega_{ng}-\omega_{pr} -\omega_{pu})t}  \Bigg].
\label{25problevel2}
\end{eqnarray}
We set $(\omega_{ng}-2\omega_{pr})t=x, \ (\omega_{ng}-\omega_{pr}-\omega_{pu})t=y, \ (\omega_{ng}-2\omega_{pu})t=z, \ E_{pr}^2/\omega_{pr}=A, \ \frac{E_{pr} E_{pu}}{\omega_{pu}}+\frac{E_{pr} E_{pu}}{\omega_{pr}}=B, \ E_{pu}^2/\omega_{pu}=C$ so that Eq. \ref{25problevel2} simplifies to:
\begin{eqnarray}
a_n^{(2)}(t)=\frac{t\mu_{mg} \mu_{nm}}{\hbar^2} \Bigg[ \frac{A(e^{ix}-1)}{x}+\frac{B(e^{iy}-1)}{y}+\frac{C(e^{iz}-1)}{z} \Bigg].
\label{26problevel2}
\end{eqnarray}
Then the probability is
\begin{eqnarray}
p_n^{(2)}(t)  =  \vert a_n^{(2)}(t) \vert ^2 &=& \frac{t^2 \vert \mu_{mg} \mu_{nm}\vert^2}{\hbar^4}\times \nonumber \\
 &\Bigg[& \frac{A^2(1-e^{ix}-e^{-ix}+1)}{x^2}+\frac{B^2(1-e^{iy}-e^{-iy}+1)}{y^2}+\frac{C^2(1-e^{iz}-e^{-iz}+1)}{z^2} \nonumber \\
&+& \frac{AB(e^{i(x-y)}-e^{ix}-e^{-iy}+1)}{xy}+\frac{AB(e^{-i(x-y)}-e^{-ix}-e^{iy}+1)}{xy}\nonumber \\
&+& \frac{AC(e^{i(x-z)}-e^{ix}-e^{-iz}+1)}{xz}+\frac{AC(e^{-i(x-z)}-e^{-ix}-e^{iz}+1)}{xz}\nonumber \\
&+& \frac{BC(e^{i(y-z)}-e^{iy}-e^{-iz}+1)}{yz}+\frac{BC(e^{-i(y-z)}-e^{-iy}-e^{iz}+1)}{yz}
\Bigg].
\label{27problevel2}
\end{eqnarray}
In the following we analyse the terms inside the square brackets of Eq. \ref{27problevel2} for long interaction times. First we introduce functions f(t) and g(t) given as:
\begin{equation}
f(t)=\frac{t^2(2-e^{ix}-e^{-ix})}{x^2}=\frac{t^22(1-cosx)}{x^2}
\label{28functionf}
\end{equation}
\begin{eqnarray}
g(t)&=&t^2[\frac{e^{i(x-y)}-e^{ix}-e^{-iy}+1}{xy}+\frac{e^{-i(x-y)}-e^{-ix}-e^{iy}+1}{xy}] \nonumber \\
&=&t^2[\frac{2+2cos(x-y)-2cosx-2cosy}{xy}].
\label{29functiong}
\end{eqnarray}
\begin{figure}
  \begin{center}
  \includegraphics[width=8.2 cm]{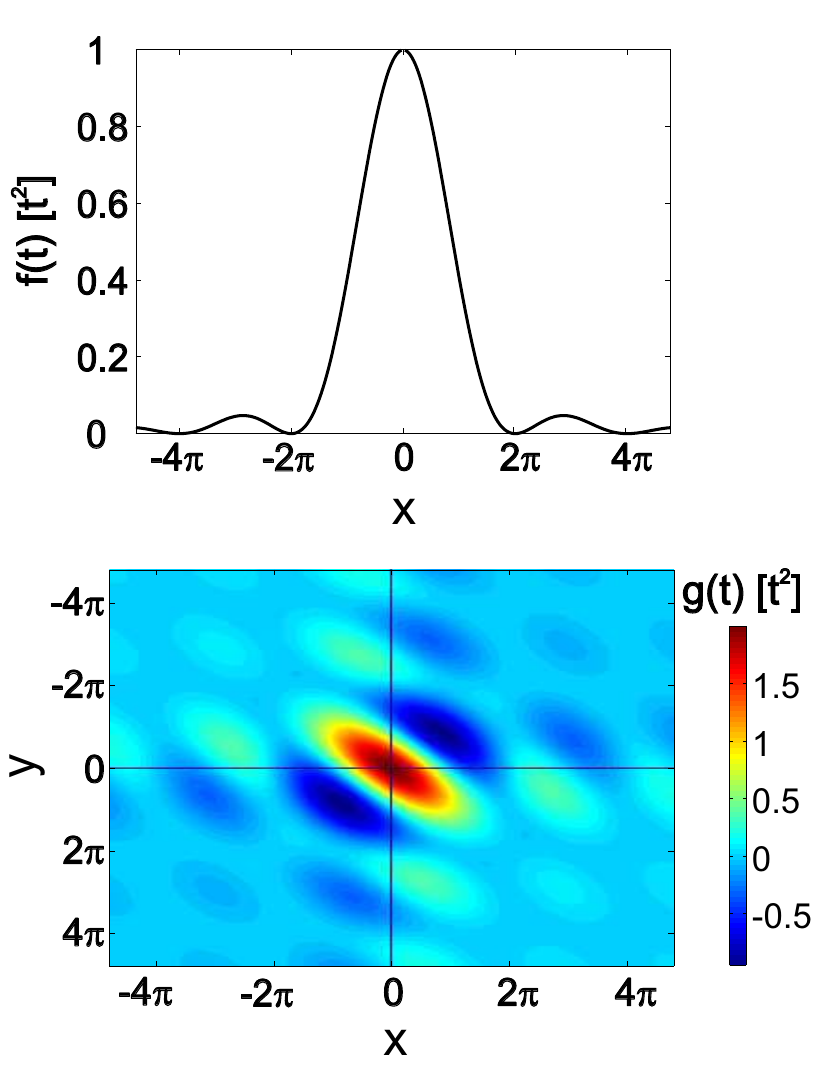}
  \caption{(color online) \emph{(a) The function f(t) versus x in Eq. \ref{28functionf}, which can be approximated as a Dirac delta function. (b) The function g(t) in Eq. \ref{29functiong}, which can be approximated as two-dimensional Dirac delta function. The absorption probability is proportional to the functions f(t) and g(t).}}
\label{yuce-f8}
  \end{center}
\end{figure}
Figure \ref{yuce-f8} (a) and (b) shows the the approximation of the functions $f(t) \ \rm{and} \ g(t)$ as Dirac delta functions for long interaction times, respectively. The peak value of $f(t)$ is $t^2$ when $x\rightarrow0$ that is $\omega_{ng}-2\omega_{pr}\rightarrow0$. The width of the central peak is of the order of $2 \pi / t$. Thus the area under the central peak is of the order of $2 \pi t$. The function f(t) can be expressed in terms of a Dirac delta function for large t as \citep{boyd.1992}:
\begin{equation}
\lim_{t\rightarrow\infty}f(t)=2\pi t \delta(\omega_{ng}-2\omega_{pr}).
\label{30limitf}
\end{equation}  
Similarly we can also write $g(t)$ in terms of Dirac delta functions for large t as:
\begin{equation}
\lim_{t\rightarrow\infty}g(t)=4\pi t \delta(\omega_{ng}-2\omega_{pr})\delta(\omega_{ng}-\omega_{pr}-\omega_{pu}).
\label{31limitg}
\end{equation}  
We can see that (Fig. \ref{yuce-f8})  $\lim_{t\rightarrow\infty} g(t)=2$, if  $\omega_{ng}-2\omega_{pr}=0 \ \rm{and} \ \omega_{ng}-\omega_{pr}-\omega_{pu}=0$ at the same time. Since $\omega_{pr}\neq\omega_{pu}$ then $g(t)=0$ for nondegenerate light sources. Later, we consider that the presence of the delta function is somewhat unphysical. Instead we use the fact that the final state n is spread into a density of final states $\rho_f(\omega_{ng})$ which is normalized such that \citep{boyd.1992}
\begin{equation}
\int_0^\infty \rho_f(\omega_{ng})d\omega_{ng}=1,
\label{32densityoflevels}
\end{equation}
and thus
\begin{equation}
\int_0^\infty \rho_f(\omega_{ng})2 \pi \delta (\omega_{ng}-2\omega_{pr})d\omega_{ng}=\rho_f(\omega_{ng}=2\omega_{pr}).
\label{33densityoflevels}
\end{equation}
Considering the fact that the first three terms inside square brackets in Eq. \ref{27problevel2} have the same form as $f(t)$, we can apply the same procedure (described from Eq. \ref{28functionf} to \ref{33densityoflevels}) for these three terms. Moreover, due to the fact that $g(t)=0$ for nondegenerate sources, the last six terms inside the square brackets in Eq. \ref{27problevel2} become also zero since they have the form of $g(t)$. As a result, the probability to be in the upper level n simplifies to
\begin{eqnarray}
p_n^{(2)}(t)&=&\frac{2\pi t \vert \mu_{mg} \mu_{nm}\vert^2}{\hbar^4} \times \nonumber \\
&[& A^2\rho_f(\omega_{ng}=2\omega_{pr})+B^2\rho_f(\omega_{ng}=\omega_{pr}+ \omega_{pu})+ C^2\rho_f(\omega_{ng}=2\omega_{pu})]. 
\label{34problevel2}
\end{eqnarray}
Since the probability for an atom to be in the upper state seems to increase linearly with time, we can define a transition rate as \citep{boyd.1992}
\begin{equation}
R_{ng}^{(2)}=\frac{p_n^{(2)}(t)}{t}.
\label{35rate2}
\end{equation}
The two-photon cross section can then be calculated via
\begin{equation}
\sigma_{ng}^{(2)}=\frac{R_{ng}^{(2)}}{I^2},
\label{36crosssection2}
\end{equation}
where $I$ is the intensity of the incident field in units of $photons/cm^2s$. We can now calculate the density of free carriers from the transition rate using
\begin{equation}
N_{eh}^{(2)}=R_{ng}^{(2)}N_{atm}\tau_{int},
\label{37carrierdensity2}
\end{equation}
where $N_{atm}$ is the number of interacting atoms per unit volume and $\tau_{int}$ the interaction time. Then we use the explicit forms of A, B, and C and we write the density of free carriers generated by two-photon absorption: 
\begin{eqnarray}
N_{eh}^{(2)} & = & {N_{atm}\tau_{int}8 \pi^{3} \vert \mu_{mg} \mu_{nm}\vert^2 \over \hbar^2 n_{0}^{2}c^2} \Bigg[ \overbrace{I_{pu}^2\rho_f(\omega_{ng}=2\omega_{pu})}^{Fig. \ref{yuce-f2}(b1)} + \overbrace{I_{pr}^2\rho_f(\omega_{ng}=2\omega_{pr})}^{Fig. \ref{yuce-f2}(b3)}  \nonumber \\
&& + \overbrace{I_{pr}I_{pu}\omega_{pr}\omega_{pu}\rho_f(\omega_{ng}=\omega_{pr}+\omega_{pu})\left(\frac{1}{\omega_{pr}^2}+\frac{2}{\omega_{pr} \omega_{pu}}+\frac{1}{\omega_{pu}^2} \right)}^{Fig. \ref{yuce-f2}(b2)} \Bigg].
 \label{39twophotonappendix}
 \end{eqnarray}
Using Eq. \ref{39twophotonappendix} we can calculate the density of free carriers generated via two-photon absorption as function of pump and probe intensities and interaction times independently. In our calculations we assume that the laser frequency is tuned to the peak of the two-photon resonance, so that $\rho_f(\omega) \approx (2 \pi \Gamma_n)^{-1}$ \citep{boyd.1992} where $\Gamma_n$ is the width of level \textit{n} (see Table~\ref{table1}). Eq. \ref{39twophotonappendix} is calculated for the general case of two-photon absorption where all permutations of pump and probe exceed the bandgap energy. One has to consider to leave out the non-resonant terms (that do not excite free carriers) to calculate the specific density of free carriers generated via two-photon absorption.

\section{Appendix C: Three-photon absorption cross section }

In order to calculate the probability amplitude $a_n^{3}$(t) for three-photon absorption (see Fig. \ref{yuce-f2}(c) for the levels) we follow the same steps as in Appendix B but modified for the three-photon absorption process. Here, we describe $\tilde{V}_{ln}$ as:
\begin{equation}
\tilde{V}_{ln}=-\mu_{ln}[E_{pr}(\omega_{pr})e^{-i \omega_{pr} t}+E_{pu}(\omega_{pu})e^{-i \omega_{pu} t}+E^*_{pr}(\omega_{pr})e^{i \omega_{pr} t}+E^*_{pu}(\omega_{pu})e^{i \omega_{pu} t}].
\label{40potential3}
\end{equation} 
We use Eq. \ref{25problevel2} and \ref{40potential3} in Eq. \ref{18schrodinger} to get $a_l^{(3)}(t)$
\begin{eqnarray}
\frac{da_l^{(3)}}{dt}&=&-(i\hbar)^{-1} \sum_{mn} a_n^{(2)} \times \tilde{V}_{ln} e^{-i\omega_{ln}t}.
\label{41schlevel3}
\end{eqnarray}
After we carry out the multiplications in Eq. \ref{41schlevel3} we use $(\omega_{lg}=\omega_{ln}+\omega_{nm}+\omega_{mg})$ for simplification and we drop the terms with $(\omega_{ln}-\omega_{pr}) \ \rm{and} \ (\omega_{ln}-\omega_{pu})$ since they give the transient response of the process and we also drop the terms describing the stimulated emission \citep{boyd.1992}. Then we integrate Eq. \ref{41schlevel3} (see Appendix B for the similar steps). We assume that both the one-photon and two-photon transitions are highly non-resonant so that  $(\omega_{mg}-\omega_{pr}\simeq\omega_{pr}) \ , \ (\omega_{mg}-\omega_{pu}\simeq\omega_{pu}) \ ,\ (\omega_{ng}-2\omega_{pr}\simeq2\omega_{pr}) \ ,\ (\omega_{ng}-2\omega_{pu}\simeq2\omega_{pu}) \ , \rm{and} \ (\omega_{ng}-\omega_{pr}-\omega_{pu}\simeq\omega_{pr}+\omega_{pu})$. Finally, we get 
\begin{eqnarray}
a_l^{(3)}(t)&=&\frac{t\mu_{ln} \mu_{nm} \mu_{mg}}{\hbar^3} \Bigg[ \frac{E_{pr}^3}{2\omega_{pr}^2} \frac{(e^{i(\omega_{lg}-3\omega_{pr})t}-1)}{(\omega_{lg}-3\omega_{pr})t}
+\frac{E_{pr}^2E_{pu}}{2\omega_{pr}^2} \frac{(e^{i(\omega_{lg}-2\omega_{pr}-\omega_{pu})t}-1)}{(\omega_{lg}-2\omega_{pr}-\omega_{pu})t} \nonumber \\
&+&\frac{E_{pr}^2E_{pu}}{\omega_{pu}(\omega_{pr}+\omega_{pu})} \frac{(e^{i(\omega_{lg}-2\omega_{pr}-\omega_{pu})t}-1)}{(\omega_{lg}-2\omega_{pr}-\omega_{pu})t}
+\frac{E_{pr}E_{pu}^2}{\omega_{pu}(\omega_{pr}+\omega_{pu})} \frac{(e^{i(\omega_{lg}-\omega_{pr}-2\omega_{pu})t}-1)}{(\omega_{lg}-\omega_{pr}-2\omega_{pu})t} \nonumber \\
&+&\frac{E_{pr}^2E_{pu}}{\omega_{pr}(\omega_{pr}+\omega_{pu})} \frac{(e^{i(\omega_{lg}-2\omega_{pr}-\omega_{pu})t}-1)}{(\omega_{lg}-2\omega_{pr}-\omega_{pu})t}
+\frac{E_{pr}E_{pu}^2}{\omega_{pr}(\omega_{pr}+\omega_{pu})} \frac{(e^{i(\omega_{lg}-\omega_{pr}-2\omega_{pu})t}-1)}{(\omega_{lg}-\omega_{pr}-2\omega_{pu})t}\nonumber \\
&+&\frac{E_{pu}^2E_{pr}}{2\omega_{pu}^2} \frac{(e^{i(\omega_{lg}-2\omega_{pu}-\omega_{pr})t}-1)}{(\omega_{lg}-2\omega_{pu}-\omega_{pr})t}
+\frac{E_{pu}^3}{2\omega_{pu}^2} \frac{(e^{i(\omega_{lg}-3\omega_{pu})t}-1)}{(\omega_{lg}-3\omega_{pu})t}
\Bigg].
\label{42problevel3}
\end{eqnarray}
In Eq. \ref{42problevel3} we set 
\begin{eqnarray}
&x=(\omega_{lg}-3\omega_{pr})t, \ y=(\omega_{lg}-2\omega_{pr}-\omega_{pu})t
\nonumber \\
&z=(\omega_{lg}-2\omega_{pu}-\omega_{pr})t, \ w=(\omega_{lg}-3\omega_{pu})t \nonumber \\
&A=\frac{E_{pr}^3}{2\omega_{pr}^2}, 
\ B=\frac{E_{pr}^2E_{pu}}{2\omega_{pr}^2}+\frac{E_{pr}^2E_{pu}}{\omega_{pu}(\omega_{pr}+\omega_{pu})}+\frac{E_{pr}^2E_{pu}}{\omega_{pr}(\omega_{pr}+\omega_{pu})}\nonumber \\
&C=\frac{E_{pu}^2E_{pr}}{2\omega_{pu}^2}+\frac{E_{pu}^2E_{pr}}{\omega_{pu}(\omega_{pr}+\omega_{pu})}+\frac{E_{pu}^2E_{pr}}{\omega_{pr}(\omega_{pr}+\omega_{pu})}, 
\ D=\frac{E_{pu}^3}{2\omega_{pu}^2}
\label{43xyz}
\end{eqnarray}
then Eq. \ref{42problevel3} simplifies to:
\begin{eqnarray}
a_l^{(3)}(t)=\frac{t\mu_{mg} \mu_{nm} \mu_{ln}}{\hbar^3} \Bigg[ \frac{A(e^{ix}-1)}{x}+\frac{B(e^{iy}-1)}{y}+\frac{C(e^{iz}-1)}{z} +\frac{D(e^{iw}-1)}{w}\Bigg].
\label{44problevel3}
\end{eqnarray}
Then the probability is
\begin{eqnarray}
p_l^{(3)}(t)  =  \vert a_l^{(3)}(t) \vert ^2 &=& \frac{t^2 \vert \mu_{mg} \mu_{nm} \mu_{ln}\vert^2}{\hbar^6}\times \nonumber \\
&\Bigg[& \frac{A^2(1-e^{ix}-e^{-ix}+1)}{x^2}
+\frac{B^2(1-e^{iy}-e^{-iy}+1)}{y^2} \nonumber \\
&+&\frac{C^2(1-e^{iz}-e^{-iz}+1)}{z^2}
+\frac{D^2(1-e^{iw}-e^{-iw}+1)}{w^2} \nonumber \\
&+& \frac{AB(e^{i(x-y)}-e^{ix}-e^{-iy}+1)}{xy}+\frac{AB(e^{-i(x-y)}-e^{-ix}-e^{iy}+1)}{xy}\nonumber \\
&+& \frac{AC(e^{i(x-z)}-e^{ix}-e^{-iz}+1)}{xz}+\frac{AC(e^{-i(x-z)}-e^{-ix}-e^{iz}+1)}{xz}\nonumber \\
&+& \frac{AD(e^{i(x-w)}-e^{ix}-e^{-iw}+1)}{xw}+\frac{AD(e^{-i(x-w)}-e^{-ix}-e^{iw}+1)}{xw}\nonumber \\
&+& \frac{BC(e^{i(y-z)}-e^{iy}-e^{-iz}+1)}{yz}+\frac{BC(e^{-i(y-z)}-e^{-iy}-e^{iz}+1)}{yz}\nonumber \\
&+& \frac{BD(e^{i(y-w)}-e^{iy}-e^{-iw}+1)}{yw}+\frac{BD(e^{-i(y-w)}-e^{-iy}-e^{iw}+1)}{yw}\nonumber \\
&+& \frac{CD(e^{i(z-w)}-e^{iz}-e^{-iw}+1)}{zw}+\frac{CD(e^{-i(z-w)}-e^{-iz}-e^{iw}+1)}{zw}
\Bigg].
\label{45problevel3}
\end{eqnarray}
If we analyze the terms inside the square brackets in Eq. \ref{45problevel3} for large interaction times, following the same steps used in Appendix B (from Eq. \ref{28functionf} to Eq. \ref{33densityoflevels}), the probability to be in upper level \textit{l} simplifies to
\begin{eqnarray}
p_l^{(3)}(t)=\frac{2\pi t \vert \mu_{mg} \mu_{nm} \mu_{ln}\vert^2}{\hbar^6} &[ A^2\rho_f(\omega_{lg}=3\omega_{pr})+B^2\rho_f(\omega_{lg}=2\omega_{pr}+ \omega_{pu})\nonumber \\
 &+C^2\rho_f(\omega_{lg}=\omega_{pr}+ 2\omega_{pu})+D^2\rho_f(\omega_{lg}=3\omega_{pu})]. 
\label{46problevel3}
\end{eqnarray}
We can define a transition rate as \citep{boyd.1992}
\begin{equation}
R_{lg}^{(3)}=\frac{p_l^{(3)}(t)}{t}.
\label{47rate3}
\end{equation}
The three-photon cross section can then be calculated via
\begin{equation}
\sigma_{lg}^{(3)}=\frac{R_{lg}^{(3)}}{I^3},
\label{48crosssection3}
\end{equation}
where $I$ is the intensity of the incident field in the units of $photons/cm^2s$. We can now calculate the density of free carriers from the transition rate using
\begin{equation}
N_{eh}^{(3)}=R_{lg}^{(3)}N_{atm}\tau_{int},
\label{49carrierdensity3}
\end{equation}
where $N_{atm}$ is the number of interacting atoms in the unit volume and $\tau_{int}$ the interaction time. Then we write the density of free carriers generated by three-photon absorption by simply inserting the functions in Eq. \ref{43xyz} into Eq. \ref{46problevel3}:
\begin{eqnarray}
N_{eh}^{(3)} & = & R_{lg}^{(3)}N_{atm}\tau_{int}={N_{atm}\tau_{int}16 \pi^{4}\vert \mu_{mg} \mu_{nm} \mu_{ln}\vert^2 \over \hbar^3 n_{0}^{3}c^3}\Bigg[
\overbrace{\frac{I_{pu}^3\rho_f(\omega_{lg}=3\omega_{pu})}{4\omega_{pu}}}^{Fig. \ref{yuce-f2}(c1)} +\overbrace{\frac{I_{pr}^3\rho_f(\omega_{lg}=3\omega_{pr})}{4\omega_{pr}}}^{Fig. \ref{yuce-f2}(c4)}  \nonumber \\
&& + \overbrace{I_{pr}I_{pu}^2\omega_{pr}\omega_{pu}^2 \rho_f(\omega_{lg}=\omega_{pr}+2\omega_{pu})\bigg(\frac{1}{4\omega_{pu}^4}+\frac{1}{\omega_{pu}^2 (\omega_{pr} +\omega_{pu})^2}+ \frac{1}{\omega_{pr}^2 (\omega_{pr}+\omega_{pu})^2} }^{Fig. \ref{yuce-f2}(c2)} \nonumber \\ 
&& + \overbrace{ \frac{1}{\omega_{pr} \omega_{pu}^2(\omega_{pr}+\omega_{pu})}+\frac{1}{\omega_{pu}^3(\omega_{pr}+\omega_{pu})}+\frac{2}{\omega_{pr}\omega_{pu}(\omega_{pr}+\omega_{pu})^2} \bigg) }^{Fig. \ref{yuce-f2}(c2)} \nonumber \\ 
&& + \overbrace{ I_{pr}^2I_{pu}\omega_{pr}^2\omega_{pu} \rho_f(\omega_{lg}=2\omega_{pr}+\omega_{pu}) \bigg(\frac{1}{4\omega_{pr}^4}+\frac{1}{\omega_{pu}^2 (\omega_{pr} +\omega_{pu})^2}  + \frac{1}{\omega_{pr}^2 (\omega_{pr}+\omega_{pu})^2} }^{Fig. \ref{yuce-f2}(c3)} \nonumber \\ 
&&+ \overbrace{ \frac{1}{\omega_{pr}^2 \omega_{pu}(\omega_{pr}+\omega_{pu})} +\frac{1}{\omega_{pr}^3(\omega_{pr}+\omega_{pu})}+\frac{2}{\omega_{pr}\omega_{pu}(\omega_{pr}+\omega_{pu})^2}\bigg) }^{Fig. \ref{yuce-f2}(c3)} \Bigg].
 \label{51threephotonappendix}
 \end{eqnarray}
Using Eq. \ref{51threephotonappendix} we are able to explicitly calculate the density of free carriers generated via three-photon absorption as function of pump and probe intensities and interaction times independently. In our calculations we assume that the laser frequency is tuned to the peak of the two-photon resonance, so that $\rho_f(\omega) \approx (2 \pi \Gamma_l)^{-1}$ \citep{boyd.1992} where $\Gamma_l$ (see Table~\ref{table1}) is the width of level \textit{l}. Eq. \ref{51threephotonappendix} is calculated for the general case of three-photon absorption where all the permutations of pump and probe exceeds the bandgap energy. One has to consider to leave out the non-resonant terms to calculate the specific density of free carriers generated via three-photon absorption.

\section*{Acknowledgments} 
We thank Allard Mosk, and Pepijn Pinkse for stimulating discussions. This research was supported by Smartmix Memphis, and the QSWITCH ANR project (to JMG). This work is also part of the research program of FOM, which is financially supported by NWO. 


\end{document}